\documentclass[prb,twocolumn,amsmath,amssymb,superscriptaddress,longbibliography]{revtex4-2}
\usepackage{color,times}
\usepackage{dcolumn}
\usepackage{multirow}
\usepackage{amssymb,amsfonts,amsmath,graphicx}
\usepackage{epsfig}
\usepackage{xcolor}
\usepackage{SIunits}
\usepackage{braket}
\usepackage{epstopdf}
\usepackage[utf8]{inputenc}
\usepackage[english]{babel}
\definecolor{darkblue}{rgb}{0, 0, 0.8}
\usepackage[colorlinks=true, breaklinks=true, linkcolor=darkblue, citecolor=darkblue, urlcolor=darkblue]{hyperref}

\graphicspath{{..//}}

\definecolor{darkgreen}{rgb}{0, 0.5, 0}

\usepackage[normalem]{ulem}

\usepackage{chngcntr}
\counterwithout{equation}{section}
\addtocounter{equation}{0}

\begin{document}

\title{Scalable spin squeezing in a dipolar Rydberg atom array}

\author{Guillaume~Bornet$^*$}
\affiliation{Universit\'e Paris-Saclay, Institut d'Optique Graduate School,\\
		CNRS, Laboratoire Charles Fabry, 91127 Palaiseau Cedex, France}

\author{Gabriel~Emperauger$^*$}
\affiliation{Universit\'e Paris-Saclay, Institut d'Optique Graduate School,\\
		CNRS, Laboratoire Charles Fabry, 91127 Palaiseau Cedex, France}		
		
\author{Cheng~Chen$^*$}
\affiliation{Universit\'e Paris-Saclay, Institut d'Optique Graduate School,\\
		CNRS, Laboratoire Charles Fabry, 91127 Palaiseau Cedex, France}

\author{Bingtian~Ye$^*$}
\affiliation{Department of Physics, Harvard University, Cambridge, Massachusetts 02138 USA}
		
\author{Maxwell~Block$^*$}
\affiliation{Department of Physics, Harvard University, Cambridge, Massachusetts 02138 USA}

\author{Marcus~Bintz}
\affiliation{Department of Physics, Harvard University, Cambridge, Massachusetts 02138 USA}

\author{Jamie A.~Boyd}
\affiliation{Universit\'e Paris-Saclay, Institut d'Optique Graduate School,\\
		CNRS, Laboratoire Charles Fabry, 91127 Palaiseau Cedex, France}

\author{Daniel~Barredo}
\affiliation{Universit\'e Paris-Saclay, Institut d'Optique Graduate School,\\
		CNRS, Laboratoire Charles Fabry, 91127 Palaiseau Cedex, France}
\affiliation{Nanomaterials and Nanotechnology Research Center (CINN-CSIC), 
		Universidad de Oviedo (UO), Principado de Asturias, 33940 El Entrego, Spain}	

\author{Tommaso~Comparin}
\affiliation{Univ Lyon, Ens de Lyon, CNRS, Laboratoire de Physique, F-69342 Lyon, France}	

\author{Fabio~Mezzacapo}
\affiliation{Univ Lyon, Ens de Lyon, CNRS, Laboratoire de Physique, F-69342 Lyon, France}	

\author{Tommaso~Roscilde}
\affiliation{Univ Lyon, Ens de Lyon, CNRS, Laboratoire de Physique, F-69342 Lyon, France}	

\author{Thierry~Lahaye}
\affiliation{Universit\'e Paris-Saclay, Institut d'Optique Graduate School,\\
		CNRS, Laboratoire Charles Fabry, 91127 Palaiseau Cedex, France}

\author{Norman~Y.~Yao}
\affiliation{Department of Physics, Harvard University, Cambridge, Massachusetts 02138 USA}
\affiliation{Department of Physics, University of California, Berkeley, California 94720 USA}
\affiliation{Materials Sciences Division, Lawrence Berkeley National Laboratory, Berkeley, CA 94720, USA}

\author{Antoine~Browaeys}
\affiliation{Universit\'e Paris-Saclay, Institut d'Optique Graduate School,\\
		CNRS, Laboratoire Charles Fabry, 91127 Palaiseau Cedex, France}

\date{\today}

\maketitle

{\bf The standard quantum limit bounds the precision of measurements that can be achieved by ensembles of uncorrelated particles. 
Fundamentally, this limit arises from the non-commuting nature of quantum mechanics, 
leading to the presence of fluctuations often referred to as quantum projection noise. 
Quantum metrology relies on the use of non-classical states of many-body systems in order 
to enhance the precision of measurements beyond the standard quantum limit~\cite{Giovannetti2011,Pezze2018}. 
To do so, one can reshape the quantum projection noise---a strategy known as \emph{squeezing}~\cite{Wineland1992,Kitagawa1993}. 
In the context of many-body spin systems, one typically utilizes all-to-all interactions 
(e.g.~the one-axis twisting model~\cite{Kitagawa1993}) between the constituents to generate the 
structured entanglement characteristic of spin squeezing~\cite{Ma2011}.
Motivated by recent theoretical work~\cite{Perlin2020,Comparin2022PRA,Comparin2022a,Comparin2022b,Block2023}, 
here we explore the prediction that short-range interactions---
and in particular, the two-dimensional dipolar XY model---can also enable the realization of scalable spin squeezing.
Working with a dipolar Rydberg quantum simulator of up to $N=100$ atoms, we demonstrate that quench 
dynamics from a polarized initial state lead to spin squeezing that improves with increasing system 
size up to a maximum of $-3.5\pm0.3$~$\textrm{dB}$ (prior to correcting for detection errors, 
or approximately $-5\pm0.3$~$\textrm{dB}$ after correction). 
Finally, we present two independent refinements: first, using a multistep spin-squeezing protocol 
allows us to further enhance the squeezing by $\sim 1$~dB, and second, leveraging 
Floquet engineering to realize Heisenberg interactions, we demonstrate the ability to 
extend the lifetime of the squeezed state by freezing its dynamics.
}

The past decade has witnessed the use of squeezed states of light and spin ensembles to 
improve upon a multitude of applications, ranging from gravitational wave detectors~\cite{Tse2019} 
and atom interferometers~\cite{Hosten2016} to optical atomic clocks~\cite{PedrozoPenafiel2020,Robinson2022}. 
The realization of spin squeezing via global interactions has been demonstrated using a variety of platforms, 
including atomic vapors coupled to light, trapped ions, ultracold gases and cavity QED~\cite{Pezze2018}. 
Whether short-range interaction (decaying as a power of the distance larger than the dimensionality) 
can yield {\it scalable} spin squeezing
has remained an essential open question~\cite{FossFeig2016,Perlin2020,Gil2014}. 
Recent theoretical advances point to an affirmative answer~\cite{Perlin2020,Comparin2022PRA,Comparin2022a,Comparin2022b,Young2022,Block2023}, 
proposing a deep connection between spin squeezing and continuous symmetry breaking (CSB)~\cite{Comparin2022PRA,Comparin2022b,Block2023,Roscilde2023}. 
This connection to CSB order broadens the landscape of systems expected 
to exhibit scalable spin squeezing, and suggests that both power-law interactions, 
and even nearest-neighbour couplings, can lead to sensitivity beyond the standard 
quantum limit~\cite{Comparin2022b,Block2023}. Of particular relevance is the ferromagnetic, 
dipolar XY model; indeed, this model is naturally realized in a number of quantum 
simulation platforms ranging from ultracold 
molecules~\cite{Yan2013,Bao2022, Holland2022,Christakisetal2023} and solid-state spin 
defects~\cite{Wolfowicz2021} to Rydberg atom arrays~\cite{Leseleuc2017, Chen2023}.

In this work, we demonstrate the generation of spin-squeezed states using a square lattice 
of up to $N = 100$~Rydberg atoms. 
Our main results are three-fold. 
First, we explore the quench dynamics of an initially polarized spin-state evolving under 
the dipolar XY model, using a procedure analogous to the one introduced
for the case of all-to-all interactions \cite{Kitagawa1993}. 
We show that the resulting state exhibits spin squeezing and characterize the generation of 
multipartite entanglement as a function of time. Moreover, the squeezing improves with increasing system size, 
providing evidence for the existence of scalable spin squeezing. Second, we introduce a multi-step approach to squeezing, 
where the quench dynamics are interspersed with microwave rotations. We demonstrate that this technique leads to an 
improvement in the amount of spin squeezing and also enables the squeezing to persist to longer time-scales. 
Finally, motivated by metrological applications, we show that it is possible to freeze the squeezing dynamics 
(e.g.~when accumulating a signal) by performing Floquet engineering. In particular, we transform the dipolar 
XY model into a dipolar Heisenberg model~\cite{Geier2021,Scholl2022}, so that the squeezing remains constant in time.

\begin{figure}[b]
\includegraphics{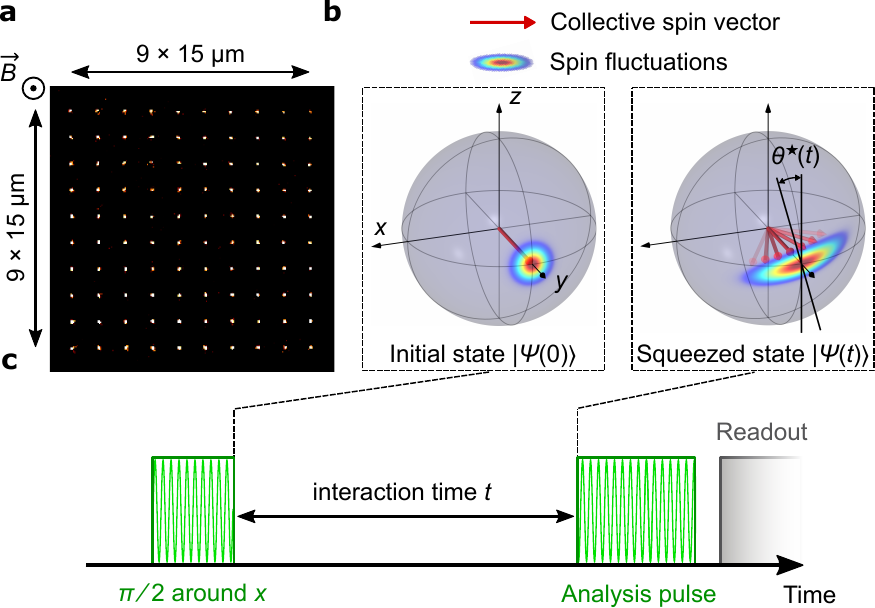}
\caption{\textbf{Generation of spin-squeezed states in a dipolar Rydberg atom array.}
\textbf{a},~Fluorescence image of a fully assembled $10\times10$ $^{87}$Rb array. 
\textbf{b},~Spin fluctuations represented via the Husimi Q-distributions (colored area)~\cite{Pezze2018} 
of the initial coherent spin state 
$\left\vert \rightarrow\cdots\rightarrow \right\rangle$ (left panel) and of a squeezed state 
obtained during the dynamics (right panel), depicted on a generalized Bloch sphere. 
The angle $\theta^\star(t)$ corresponds to the direction of the narrowest 
noise distribution.
The squeezed state is schematically depicted by a superposition of coherent states (red arrows). 
\textbf{c},~The sequence of microwave pulses corresponding to the spin squeezing protocol. 
A first $\pi/2$ pulse initializes all the spins along $\hat{y}$. 
By tuning the duration and phase of a second (analysis) microwave pulse prior to readout, one can rotate the spin distribution around 
$\hat{y}$ to measure the variance $\text{Var}(J_{\theta})$ along any direction $\theta$, or around 
$\hat{x}$ to measure the spin length $|\langle J_y \rangle |$.
}
\label{fig:fig1}
\end{figure}

\begin{figure*}[t]
\includegraphics{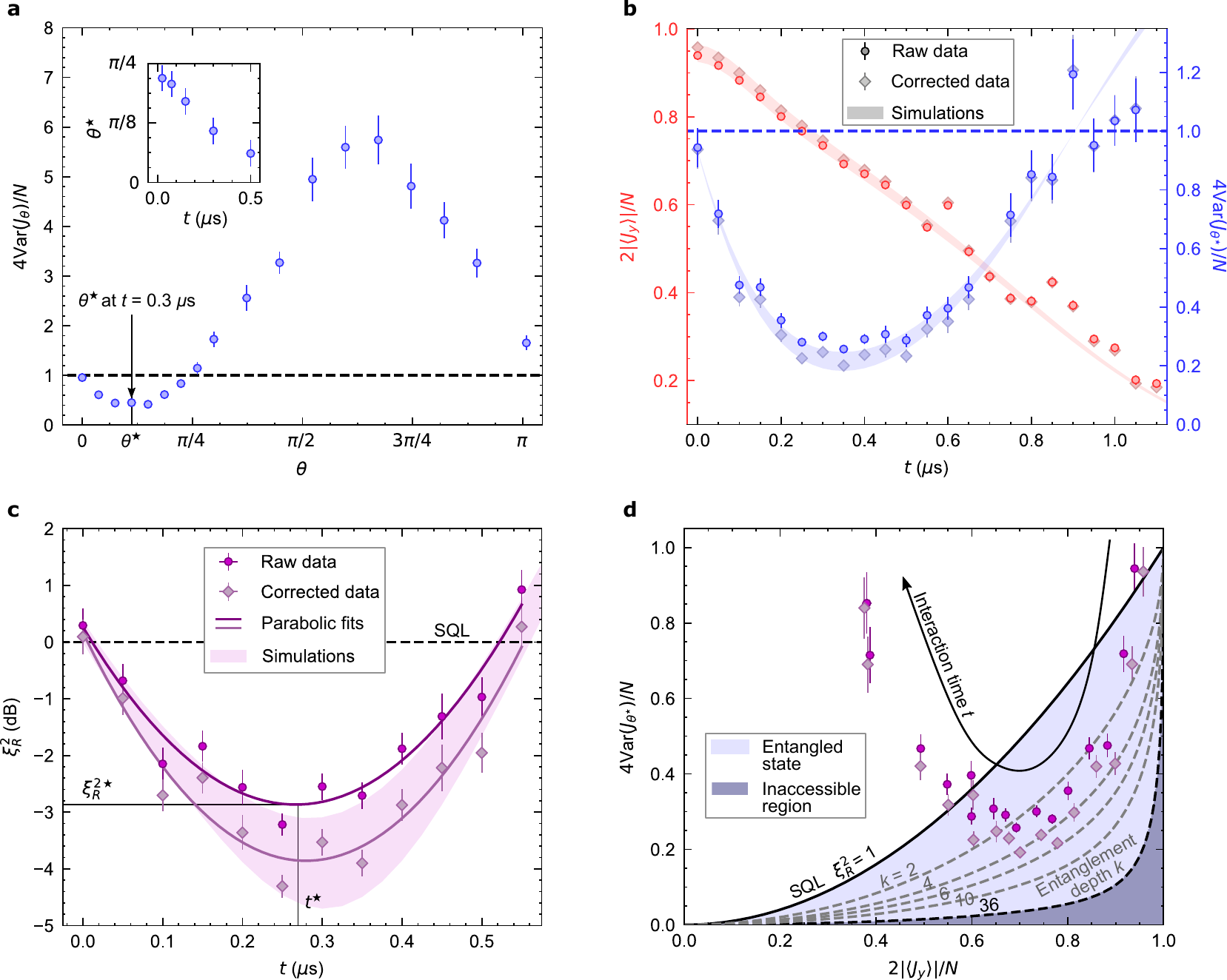}
\caption{\textbf{Dynamical evolution of spin squeezing in an $N = 6\times6$ array.}
\textbf{a},~Determination of the angle $\theta^{\star}$ that minimizes the spin fluctuations for a fixed interaction time, $t = 0.3~\mu$s. 
The inset shows $\theta^{\star}(t)$ determined for different times, $t$. 
The dashed line in \textbf{a} and \textbf{b} corresponds to the uncorrelated case $4\text{Var}(J_{\theta})/N=1$.
\textbf{b},~Measurements of the spin length $|\langle J_y \rangle|$ (red circles) and of the minimum variance 
$\text{Var}(J_{\theta^{\star}})$ (blue circles). The diamond markers are the data 
corrected for the detection errors, as described in the Methods.
The shaded regions represent the results of the unitary spin dynamics, without any free parameter, 
including $97.5 \pm 1\%$ ($99 \pm 1\%$) detection efficiency for $\ket{\uparrow}$ ($\ket{\downarrow}$).
\textbf{c},~Squeezing parameter $\xi_R^2(t)$ as a function of time. The solid curves are parabolic fits used to determine the optimal squeezing parameter $\xi^{2\star}_R$ and the optimal squeezing time $t^\star$. As in \textbf{c}, the shaded area shows simulations including $\pm 1\%$ uncertainty in the detection efficiencies.
\textbf{d}, Parametric plot of the variance as a function of the spin length. The colored area, delimited by the black solid curve $\xi_R^2 = 1$, depicts the region where entanglement exists in the system. The grey dashed curves correspond to entanglement depths of $k$ and the dashed black curve to a maximal entanglement depth of $k = 36$. The black arrow shows the direction of increasing interaction time. 
}
\label{fig:fig2}
\end{figure*}

Our experimental setup~\cite{Scholl2021} consists of a two-dimensional square array of $^{87}$Rb atoms 
trapped in optical tweezers (see Fig.~\ref{fig:fig1}a). 
To implement the dipolar XY model~\cite{Browaeys2020}, we rely on resonant dipole-dipole interactions 
between two Rydberg states of opposite parities. In particular, we encode an effective spin-$1/2$ 
degree of freedom as $\left|\uparrow\right\rangle = |60S_{1/2}, m_j = +1/2\rangle$ and 
$\left|\downarrow\right\rangle = |60P_{3/2}, m_j = -1/2\rangle$, leading to an interaction Hamiltonian:
\begin{equation}\label{Eq:1}
H_{\rm XY}= - {\frac{J}{2}}
\sum_{i < j} \frac{a^3}{r_{ij}^3} (\sigma^x_i \sigma^x_j + \sigma^y_i \sigma^y_j),
\end{equation}
where $\sigma_i^{x,y,z}$ are Pauli matrices, $r_{ij}$ is the distance between spins $i$ and $j$, $J/h= 0.25~$MHz 
is the dipolar interaction strength, and $a=15~\mu$m is the lattice spacing. A magnetic field perpendicular to the lattice 
plane defines the quantization axis and ensures that the dipolar interactions are isotropic.

We begin by investigating the squeezing dynamics generated by $H_{\rm XY}$.  
The atoms are initially excited from the ground state to the Rydberg state $\ket{\uparrow}$, 
using stimulated Raman adiabatic passage (see Methods). Using a microwave $\pi/2$-pulse 
tuned to the transition between the spin states, 
we prepare an initial coherent spin state along the $y$-axis, 
$\left\vert \psi (0) \right\rangle = \left | \rightarrow \cdots \rightarrow \right \rangle$ (see Fig.~\ref{fig:fig1}b). 
Next, we allow the system to evolve under $H_{\rm XY}$ and measure the squeezing as a function of time.

Since squeezing manifests as a change in the shape of the noise distribution, 
one must measure the variance of the collective spin operator in the plane perpendicular 
to the mean spin direction; to this end, we define $J_{\theta} = \cos(\theta)J_z + \sin(\theta)J_x$, where $J_{x,y,z} = \frac{1}{2}\sum_{i}\sigma^{x,y,z}_i$ are collective spin operators. We characterize the amount of spin squeezing via the parameter~\cite{Wineland1992,Wineland1994}, 
\begin{equation}\label{Eq:2}
\xi_R^2(t) = \frac{N\min_{\theta} \left( \text{Var}\left(J_{\theta}\right) \right)}{{\langle J_y \rangle}^2},
\end{equation}
which quantifies the metrological gain in a Ramsey interferometry experiment. To measure $|\langle J_y \rangle|$, we simply rotate the state $|\psi(t)\rangle$ back to the $z$-axis using a second $\pi/2$-pulse around $x$. To measure $\text{Var}\left(J_{\theta}\right)$, we instead perform a microwave rotation around the $y$-axis, where the angle $\theta$ is tuned via the duration of the pulse. Finally, we read out the state of each atom with a detection efficiency of $97.5\%$ for $\ket{\uparrow}$ and $99\%$ for $\ket{\downarrow}$ (see Methods). Operationally, each experimental sequence is repeated $\sim200$ times, and from this series of snapshots, we calculate the average and variance of all collective spin operators. For a given interaction time, $t$, the noise distribution has a specific direction of smallest uncertainty, corresponding to the angle $\theta^{\star} (t)$ that minimizes the variance of $J_{\theta}$ (see Fig.~\ref{fig:fig1}b). Beginning with a 6$\times$6 array, we measure $\text{Var}\left(J_{\theta}\right)$ as a function of $\theta$ for $t=0.3~\mu$s. As shown in Fig.~\ref{fig:fig2}a, the variance exhibits a sinusoidal shape that reveals the underlying elliptical distribution of the spin fluctuations and allows us to determine $\theta^{\star}$. We then investigate the time evolution of both $|\langle J_y \rangle|$ and $\text{Var}\left(J_{\theta^\star}\right)$. As the system evolves, the initial coherent spin state expands into a superposition of states (fan of red arrows, Fig.~\ref{fig:fig1}b), which causes the mean spin length, $|\langle J_y \rangle|$ (red circles, Fig.~\ref{fig:fig2}b), to decay toward zero~\cite{Geier2021, Scholl2022}. At the same time, the variance of $J_{\theta}$ (blue circles, Fig.~\ref{fig:fig2}b) initially decreases below its $t=0$ value, reaches a minimum, and then increases, exceeding its $t=0$ value at late times~\cite{Comparin2022b}.

Taken together, $|\langle J_y \rangle|$ and $\text{Var}\left(J_{\theta^\star}\right)$ allow us to reconstruct the squeezing parameter $\xi_R^2$ (or $10\log_{10}(\xi_R^2)$ when expressed in dB) as a function of time. As illustrated in Fig.~\ref{fig:fig2}b, the dynamics of the squeezing parameter are qualitatively similar to those of the variance: $\xi_R^2$ initially decreases below the standard quantum limit (SQL), reaches an optimum $\xi_R^{2\star}$ at time $t^{\star}$, and then increases, exceeding the SQL at late times. The system remains in a squeezed state (i.e.~$\xi_R^2<1$) for approximately $0.5~\mu$s and exhibits an optimal squeezing parameter of $-2.7\pm0.3~$dB. The optimal squeezing is highly sensitive to detections errors, and analytically correcting for these errors (diamond markers, Fig.~\ref{fig:fig2}, see Methods) leads to a minimum squeezing parameter of $-3.9\pm0.3~$dB. However, even this corrected value does not reach the optimum (approximately $-6.7~$dB) predicted for the dipolar XY model. We attribute this to two other types of experimental imperfections, which also degrade the squeezing parameter: errors in the initial state preparation and imperfections in our microwave pulses. In contrast to detection errors, these imperfections directly affect the many-body squeezing dynamics (see Methods); accounting for these additional errors leads to significantly better agreement between theory and experiment (Fig.~\ref{fig:fig2}c). 

At a fundamental level, a squeezing parameter $\xi_R^2<1$ necessarily implies 
the presence of entanglement in our system~\cite{Sorensen2001b,Sorensen2001}. 
We quantify the entanglement depth as a function of time; 
an entanglement depth of $k$ means that the many-body state cannot 
be written as a statistical mixture of states factorized into clusters containing 
up to $(k-1)$ particles -- that is, at least one $k$-particle subsystem is entangled
~\cite{Sorensen2001,Esteve2008,Riedel2010}. 
For a particular spin length, $|\langle J_y \rangle|$, the minimum attainable variance of the quantum state gives 
a lower bound on the entanglement depth~\cite{Sorensen2001}. Fixed contours of this bound for different values of 
$k$ are shown in Fig.~\ref{fig:fig2}d:
if a data point falls below the line labelled by $k$, the entanglement depth is thus at least $k+1$.
The many-body dynamics of our system leads to a state whose entanglement depth increases rapidly at early times. 
Near the optimal squeezing time, $t^{\star}$, the entanglement depth reaches a maximum of $k = 3$ 
(for the measurement-corrected data, we find $k = 5$) for our 36-atom system. \\

\begin{figure*}
\includegraphics{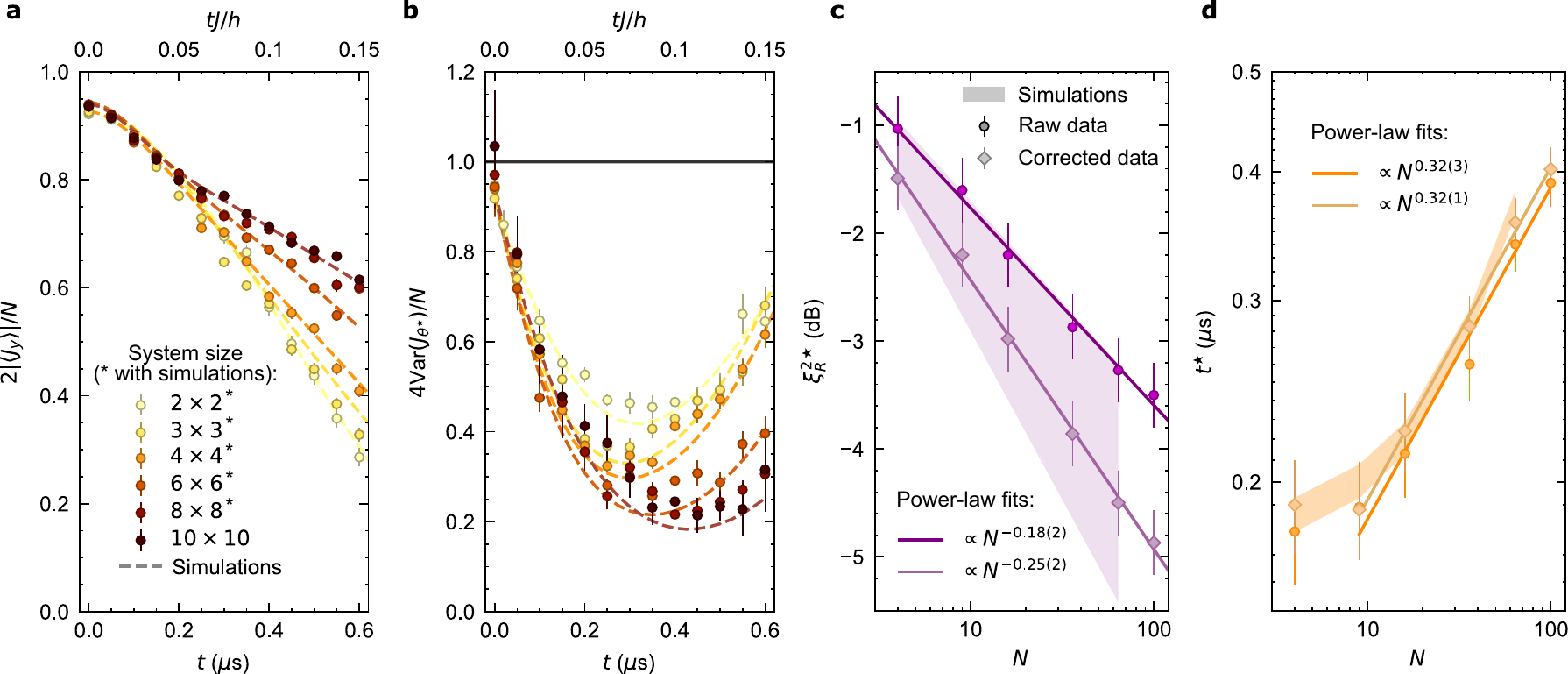}
\caption{\textbf{Scalable spin squeezing in the two-dimensional, ferromagnetic, dipolar XY model.}
\textbf{a}, \textbf{b}, Measurement of the spin length $|\langle J_y \rangle|$ and of $\text{Var}(J_{\theta^{\star}})$ as 
a function of time for various system sizes. The dashed lines correspond to the results of matrix-product state simulations 
without any adjustable parameters, which are limited to system sizes of $8 \times 8$ (see Methods). 
\textbf{c}, \textbf{d}, Minimum squeezing parameter $\xi_{R}^{2\star}$ and associated optimal interaction 
time $t^{\star}$, as a function of $N$. The circles and diamonds correspond to the raw and detection-error 
corrected data, respectively. The solid lines are power-law fits. The shaded regions  are the results of the simulations 
for values of the detection efficiency of $\ket{\uparrow}$ ($\ket{\downarrow}$),  $97.5 \pm 1\%$ ($99 \pm 1\%$), 
between their lower and upper limit.
}
\label{fig:fig3}
\end{figure*}

One of the distinguishing features of spin squeezing in all-to-all interacting models 
is that it is \emph{scalable}---the optimal squeezing parameter, $\xi_R^{2\star}$, 
scales non-trivially with system size as $N^{-\nu}$ with $\nu = 2/3$ \cite{Kitagawa1993}.
Whether this is the case for power-law interacting systems is significantly more subtle. 
A heuristic way to understand the emergence of early time squeezing dynamics in the dipolar XY model consists of rewriting the interaction as: $(\sigma^x_i \sigma^x_j + \sigma^y_i \sigma^y_j)/r_{ij}^3 
= (\vec{\sigma_i} \cdot \vec{\sigma_j} - \sigma^z_i \sigma^z_j)/r_{ij}^3$; 
for our initial coherent spin state, the Heisenberg term yields no dynamics, while
the $\sigma^z_i \sigma^z_j/r_{ij}^3$ term approximates the squeezing dynamics generated by the all-to-all-coupled, one-axis twisting (OAT) model $H_{\textrm{OAT}} \propto \sum_{i,j} \sigma^z_i \sigma^z_j \propto J_z^2$, at short times~\cite{FossFeig2016,Roscilde2021}. 
However, this description breaks down at an $\mathcal{O}(h/J)$ time-scale 
(i.e.~as soon as the state is no longer fully polarized) and thus cannot explain the emergence of scalable spin squeezing.
Going beyond this heuristic rewriting, more rigorous arguments can be made for the emergence of scalable spin squeezing in the dipolar XY Hamiltonian (see Methods).

In particular, in power-law interacting systems, scalable spin squeezing has been recently conjectured to be closely related to continuous symmetry breaking (ferromagnetic XY) order~\cite{Perlin2020,Comparin2022PRA,Comparin2022b,Block2023}.
The mean-spin direction is the order parameter of such a system, and thus, in the ordered phase, it should equilibrate to some nonzero value; this is clearly a pre-requisite for scalable squeezing, since the denominator of the squeezing parameter, $\xi^2_R$, is precisely the square of the mean-spin length, ${\langle J_y \rangle}^2$ (Eq.~\ref{Eq:2}). 
More subtly, the low-energy spectrum associated with ferromagnetic XY order is expected to consist of so-called ``Anderson towers'', wherein the energy is proportional to $J_z^2$ (see Methods)~\cite{anderson_approximate_1952,anderson_basic_1997}.
Crucially, this leads to the emergence and persistence of OAT-like dynamics \emph{even} until late times, 
$t\sim\mathcal{O}(N h/J)$; these dynamics ``twist'' the initial quantum fluctuations, shrinking the minimum variance 
in the $y$-$z$ plane (Eq.~\ref{Eq:2}), thus leading to scalable spin squeezing.
Finally, let us emphasize that even this picture is only approximate: 
The eventual thermalization of the dipolar XY model implies that its dynamics (even at low energies) 
cannot be perfectly captured by one-axis twisting~\cite{Block2023}. 

For the dipolar XY interactions that we investigate here, continuous symmetry breaking, and thus scalable squeezing, 
is expected in $d=2$ \cite{Perlin2020,Comparin2022PRA,Comparin2022b,Block2023}, 
but not in $d=1$ \cite{Comparin2022PRA,Block2023}. 
To this end, we measure the squeezing dynamics in systems ranging from 
$N = 2\times2$ to $10\times10$ atoms. In principle, determining the minimum squeezing 
parameter requires optimizing over both time and $\theta$ for each system size; 
as $N$ increases, the optimal time, $t^{\star}$, is expected to increase while the optimal 
$\theta^{\star}$ is expected to decrease. Analogous to our previous procedure, we begin 
by extracting $\theta^{\star}$ at a fixed time $t$, and measuring the time evolution of 
$|\langle J_y \rangle|$ and $\text{Var}(J_{\theta^{\star}})$; the time at which the variance 
is minimized provides a self-consistent way to experimentally verify that we are 
working near the two-parameter optimum.

As depicted in  Fig.~\ref{fig:fig3}(a), the dynamics of $|\langle J_y \rangle|$ 
at short times ($t<0.25~\mu$s)
collapse (i.e. exhibit a size-independent decay) for all $N$ owing to rapid local relaxation of 
the magnetization (which can be analyzed using spin-wave theory, see Methods and~\cite{Roscilde2023b}). 
At later times, $|\langle J_y \rangle|$ decreases more slowly for increasing system size, 
indicative of continuous symmetry breaking order. 
The dynamics of the variance also depend on $N$ (see Fig.~\ref{fig:fig3}b): 
the minimum variance improves and occurs at later times as the system size increases. 
From these measurements, for each system size, we compute the squeezing dynamics, 
and extract both the optimal squeezing parameter, $\xi_R^{2\star}$, 
and the corresponding optimal interaction time, $t^{\star}$.

As previously mentioned, in the all-to-all interacting case, both optima are 
expected to scale with system size~\cite{Kitagawa1993, Pezze2009}. 
Recent theoretical work predicts that scalable squeezing can also arise 
in our 2D dipolar XY model \cite{Comparin2022b,Block2023,Roscilde2023}. 
This expectation is indeed borne out by our data.
As shown in  Fig.~\ref{fig:fig3}(c,d), we find that $\xi_R^{2\star} \sim N^{-\nu}$ and 
$t^{\star} \sim N^\mu$ with $\nu = 0.18(2)$ and $\mu = 0.32(3)$; 
when correcting for detection errors, we find that $\nu = 0.25(5)$, while $\mu$ does not change. 
Interestingly, the exponent that we observe for the optimal squeezing time is in agreement 
with that observed in the all-to-all coupled case, where $t^{\star} \sim N^{^{1/3}}$~\cite{Kitagawa1993}. 
However, the scaling of the optimal squeezing parameter is significantly weaker 
than that predicted for both all-to-all interactions, as well as the dipolar XY 
model~\cite{Comparin2022b,Block2023}. Again, we attribute this to a combination of 
experimental imperfections, which, when accounted for, leads to a relatively good agreement 
between theory and experiment as shown in  Fig.~\ref{fig:fig3}b and~\ref{fig:fig3}c. 
We note that this difference in agreement for $t^{\star}$ and $\xi_R^{2\star}$ 
is perhaps not unexpected; for example, measurement errors decrease the 
amount of achievable spin squeezing but do not change the optimal squeezing time.\\

As schematically depicted in  Fig.~\ref{fig:fig1}b, the fact that squeezing 
exhibits an optimum in time arises from a competition between the generation of entanglement 
and the curvature of the Bloch sphere. Microscopically, the squeezing 
dynamics causes the coherent superposition of states to wrap around the Bloch sphere, 
but squeezing (Eq.~\ref{Eq:2}) is measured via the variance projected in the 
plane perpendicular to the mean spin direction. Thus, the curvature of the Bloch sphere 
leads to a noise distribution which deviates from an elliptical shape \cite{Pezze2018} 
and manifests as additional variance (see Methods). 
This suggests that one can improve the optimum squeezing by utilizing a time-dependent protocol.
In particular, by rotating the elliptical noise distribution toward the equator, 
one can minimize the impact of the projection on the measured variance~\cite{Muessel2015, Sorelli2019}.

\begin{figure}
\includegraphics{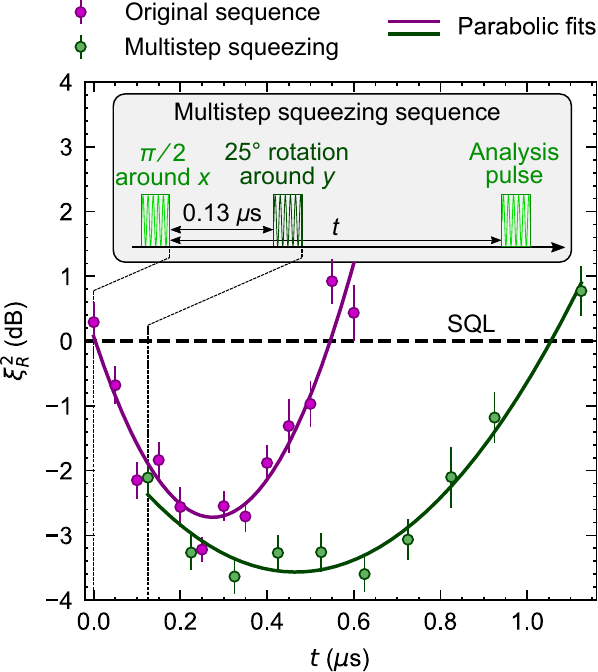}
\caption{\textbf{Multi-step spin squeezing protocol.}
Measurements of the squeezing parameter obtained with two different spin squeezing protocols for a $6\times6$ array. 
The first one (purple dots) is the original sequence illustrated in  Fig.~\ref{fig:fig1}\textbf{c}. 
The second one (dark green dots) is a multi-step sequence depicted in the inset, 
where an additional $25 ^\circ$ rotation pulse is used to rotate the elliptical noise distribution toward the equator.
The solid curves are parabolic fits to guide the eye. 
}
\label{fig:fig4}
\end{figure}

To this end, working with a $6\times6$ array, we implement a discretized, single-step version of this protocol. 
We initialize the system in the same initial state, 
$\left\vert \psi (0) \right\rangle = \left | \rightarrow \cdots \rightarrow \right \rangle$, 
and let the squeezing dynamics proceed for $t = 0.13~\mu$s. 
Then, we perform a $25^{\circ}$ rotation around the $y$-axis in order to 
nearly align the noise distribution's major axis parallel to the equator (see Methods). 
The subsequent dynamics of the squeezing parameter are shown in  Fig.~\ref{fig:fig4} (green data). 
Three effects are observed. First, the optimal squeezing occurs at a later time, 
$t^{\star} \sim 0.45~\mu$s. Second, consistent with the intuition above, 
the system remains near its optimal squeezing value for approximately twice as long. 
Third, the value of the optimal squeezing parameter is improved by approximately $1$~dB, reaching a value of $-3.6\pm0.3$~dB.\\

In order to perform sensing, it is desirable to freeze the squeezing dynamics 
while acquiring the signal of interest. The simplest way to do so is to turn off the Hamiltonian. 
However, it is challenging to directly turn off the dipolar exchange interaction between 
the Rydberg atoms.

To this end, we utilize an alternate approach, where Floquet driving~\cite{Geier2021,Scholl2022} 
engineers an effective dipolar Heisenberg interaction, 
$H_\mathrm{Heis} = - {\frac{2J}{3}} \sum_{i < j} \frac{a^3}{r_{ij}^3} \vec{\sigma}_i \cdot \vec{\sigma}_j $, 
from our original XY model. 
Crucially, the Heisenberg interaction commutes with all collective spin operators, 
ensuring that: (i) it does not change the spin squeezing and (ii) it does not affect the sensing signal associated, for example,
to the presence of a uniform external field~\footnote{The Floquet sequence that generates the Heisenberg 
interaction leads to a rescaling of the strength an external field by a factor of $1/3$.}.

\begin{figure}
\includegraphics{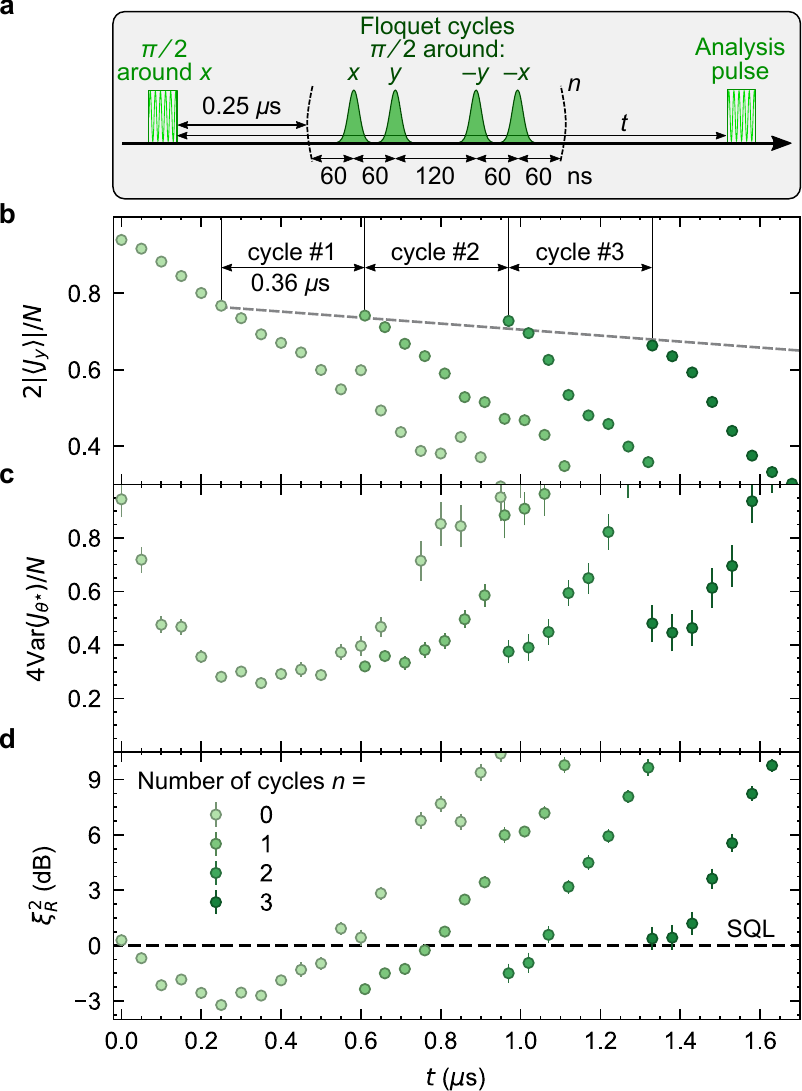}
\caption{\textbf{Floquet engineering to freeze spin squeezing.}
\textbf{a}, Experimental WAHUHA sequence using Floquet engineering to realize an effective dipolar 
Heisenberg interaction in a $N=6\times6$ array. 
The system is periodically driven using $n$ Floquet cycles, each composed of four 
$\pi/2$ Gaussian microwave pulses (of half-width $6.5~$ns at $1/\sqrt{e}$), 
whose phases are chosen to realize rotations around the $(x,y,-y,-x)$ axes.
\textbf{b}, \textbf{c}, and \textbf{d}, Spin length $|\langle J_y \rangle|$, 
minimal variance $\text{Var}(J_{\theta^{\star}})$, and squeezing parameter $\xi_R^{2}$ 
as a function of the total interaction time $t$ for different numbers, $n$, of applied Floquet cycles. 
The grey dashed line in \textbf{b} is a guide to the eye to highlight the spin length 
measured immediately after each Floquet cycle.
}
\label{fig:fig5}
\end{figure}

To explore this behavior, we let our system evolve to the optimal squeezing time and 
then attempt to freeze the dynamics via the Floquet WAHUHA sequence shown in Fig.~\ref{fig:fig5}a \cite{Waugh1968}. 
A full Floquet cycle lasts $t_\mathrm{F} = 0.36~\mu$s and for rapid driving, $J t_\mathrm{F} \ll 2\pi$, 
the time-averaged Hamiltonian is approximately $H_\mathrm{Heis}$~\cite{Scholl2022}. 
We repeat this experiment for different numbers of Floquet cycles ranging from $n = 0\text{-}3$. 
The Floquet dynamics of $|\langle J_y \rangle|$ and $\text{Var}(J_{\theta^{\star}})$ are illustrated in 
Fig.~\ref{fig:fig5}b,c. For perfectly frozen dynamics, each set of curves (with different $n$) would 
simply be off-set in time from one another. This expectation is in good agreement with the data. 
Indeed, as depicted in  Fig.~\ref{fig:fig5}b, we observe that the dynamics of $|\langle J_y \rangle|$ 
are translated in time, except for a small downward drift (indicated by the grey dashed line). 
We note that this downward drift is significantly weaker than the intrinsic dynamics of $|\langle J_y \rangle|$. 
Comparable behaviour is observed for $\text{Var}(J_{\theta^{\star}})$ [ Fig.~\ref{fig:fig5}c]. 
Finally, as illustrated by the squeezing parameter in  Fig.~\ref{fig:fig5}d, the Floquet sequence prolongs 
the time-scale over which squeezing remains below the SQL by nearly a factor three.\\

To conclude, our work represents the first observation of scalable spin squeezing in a 
many-body system with short-range, power-law interactions. 
It is complementary to the recent results obtained with Rydberg-dressed atoms \cite{Eckner2023,hines2023}
and long-range interactions in an ion string~\cite{Franke2023}.
Our findings and methods are applicable to any quantum systems 
implementing the dipolar XY Hamiltonian, such 
as molecules~\cite{Yan2013,Bao2022, Holland2022,Christakisetal2023} 
or solid-state spin defects~\cite{Wolfowicz2021}.
Within the context of tweezer arrays, our work lays the foundation for several research directions.
First, by generalizing our approach to alkaline-earth Rydberg tweezer arrays~\cite{Norcia2018,Cooper2018,Saskin2019}, 
it may be possible to map the spin squeezing from the Rydberg manifold to the so-called clock transition, 
in order to improve tweezer-based atomic clocks~\cite{Madjarov2019,Young2020}.
Second, by investigating squeezing as a function of the initial polarization, e.g. by introducing disorder in the initial state preparation, it may be possible to test theoretical predictions that spin squeezing in short-range interacting systems is fundamentally distinct from that achieved in all-to-all coupled systems~\cite{Block2023}. 
Finally, by implementing a continuous version of the multi-step squeezing protocol, it may be possible to improve the scaling of the observed spin squeezing toward the Heisenberg limit.

\begin{acknowledgments}

We acknowledge the insights of and discussions with M. P. Zaletel, B.~Halperin, B.~Roberts, C.~Laumann, E.~Davis, S.~Chern, W.~Wu, Z.~Wang, A.-M. Rey, F.~Yang and Q.~Liu.
This work is supported by
the Agence Nationale de la Recherche (ANR, project RYBOTIN and ANR-22-PETQ-0004 France 2030, project QuBitAF), 
and the European Research Council (Advanced grant No. 101018511-ATARAXIA). 
B.~Y. acknowledges support from the AFOSR MURI program (W911NF-20-1-0136).
M.~Block acknowledges support through the Department of Defense (DoD) through the National Defense Science and Engineering Graduate (NDSEG) Fellowship Program.
M.~Bintz acknowledges support from the Army Research Office (W911NF-21-1-0262).
N.~Y.~Y. acknowledges support from the U.S. Department of Energy, Office of Science, National Quantum Information Science Research Centers, Quantum Systems Accelerator (QSA).
DB acknowledges support from MCIN/AEI/10.13039/501100011033 
(RYC2018- 025348-I, PID2020-119667GA-I00, and European Union NextGenerationEU PRTR-C17.I1).
The computational results presented were performed in part using the FASRC Cannon cluster supported by the FAS Division of Science Research Computing Group at Harvard University, the Savio computational cluster resource provided by the Berkeley Research Computing programme at the University of California, and the PSMN cluster at the ENS Lyon.

\end{acknowledgments}

\section*{Author contributions}
$^*$G.B., G.E., C.C., B.Y. and M.~Block contributed equally to this work.
G.B., G.E., C.C., J.A.~B. and D.B. carried out the experiments. 
B.Y., M.~Block, M.~Bintz, T.C. and F.M. conducted the theoretical analysis and simulations. 
T.R., T.L., N.Y.Y. and A.B. supervised the work.
All authors contributed to the data analysis, progression of the project, 
and on both the experimental and theoretical side. 
All authors contributed to the writing of the manuscript.

\section*{Correspondence and requests for materials}
Correspondence and requests for materials should be addressed to Cheng Chen: cheng.chen@institutoptique.fr

\section*{Ethics Declaration}
A.B. and T.L. are co-founders and shareholders of PASQAL. The remaining authors declare no competing interests.

\bibliography{reference_squeezing}

\clearpage

\setcounter{figure}{0}
\renewcommand\thefigure{A\arabic{figure}} 

\appendix

\section*{Methods}

\subsection{Experimental methods}\label{SM:Exp_details}

\begin{figure*}[t!]
\includegraphics{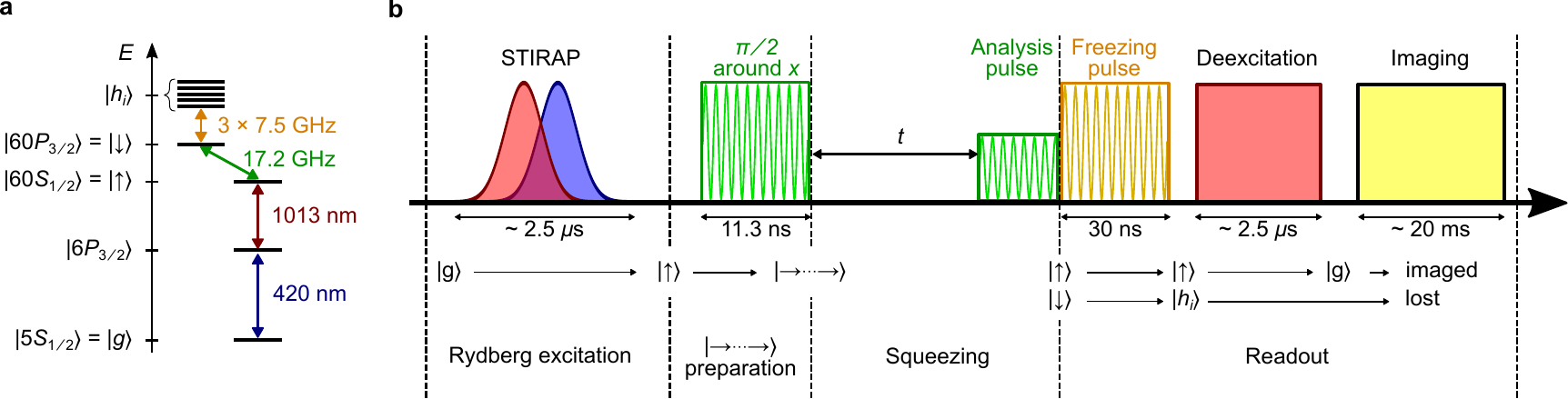}
\caption{\textbf{Experimental sequence.}
\textbf{a}, Schematics of the atomic levels relevant for the experiment.
\textbf{b}, Sequence of optical and microwave pulses (not to scale) used for all the experiments reported 
in Figs.\,\ref{fig:fig1},\ref{fig:fig2} and \ref{fig:fig3} of the main text.}
\label{fig:ExpSequ_SM}
\end{figure*}

The realization of the dipolar XY Hamiltonian relies on the $^{87}\text{Rb}$ Rydberg-atom tweezer array platform described in previous works~\cite{Barredo2016,Scholl2021}. We encode our pseudo spin states as $\ket{\uparrow} = \ket{60S_{1/2}, m_J = +1/2}$ and $\ket{\downarrow} = \ket{60P_{3/2}, m_J = -1/2}$, and couple them by using microwaves at 17.2~GHz (see Fig.~\ref{fig:ExpSequ_SM}a). The microwave field is emitted by an antenna placed outside the vacuum chamber, leading to poor control over the polarization, due to the presence of metallic parts surrounding the atoms. To isolate the $\ket{\uparrow}-\ket{\downarrow}$ transition from irrelevant Zeeman sublevels we apply a $\sim45$-G quantization magnetic field perpendicular to the array.\\

\emph{Experimental sequence} -- Figure~\ref{fig:ExpSequ_SM}(b) shows the details of the full experimental sequence. After randomly loading atoms into 1-mK deep optical tweezers (with a typical filling fraction of $60\%$), the array is assembled one atom at a time~\cite{Barredo2016}. The atoms are then cooled to a temperature of $10\,\mu$K using Raman sideband cooling and optically pumped to $\ket{g}=\ket{5S_{1/2}, F = 2, m_F = 2}$. Following this, the power of the trapping light is adiabatically ramped down reducing the tweezer depth by a factor $\sim 50$. Then, the tweezers are switched off and the atoms are excited to the Rydberg state $\ket{\uparrow}$. The excitation is performed by applying a two-photon stimulated Raman adiabatic passage (STIRAP) with 421-nm and 1013-nm lasers.

To generate $\left\vert \psi(0) \right\rangle \equiv \ket{\rightarrow \cdots \rightarrow}$, 
we first apply a global resonant microwave $\pi/2$ pulse around $x$, with a Rabi frequency
$\Omega = 2\pi \times 22.2$~MHz. After an interaction time $t$, an analysis microwave 
pulse is applied to change the measurement basis.
When measuring the $J_{\theta}$ variance, the Rabi frequency of the analysis pulse is reduced 
down to $2\pi \times 4.1$~MHz in order to perform rotations with a higher angular resolution. 
We note that the $\ket{\uparrow}-\ket{\downarrow}$ transition frequency changes slightly when 
varying the microwave Rabi frequency. We attribute this to a light shift induced by couplings between 
the other components of the microwave polarization and the Zeeman sublevels of the 
$60S_{1/2}$ and $60P_{3/2}$ manifolds. We experimentally compensate for this effect by 
detuning the microwave (e.g. for $\Omega = 2\pi \times 22.2$~MHz, the corresponding detuning is $2\pi \times 3.5$~MHz).

The experimental sequence (including detection, detailed below) 
is typically repeated for $\sim200$ defect-free assembled arrays. 
This allows us to calculate the magnetization, spin correlations and 
variance by averaging over these realizations.\\

\emph{State-detection procedure} -- The detection protocol comprises three steps.
In the first step, a $7.5~$GHz microwave pulse (i.e. the ``freezing pulse'' in 
Fig.~\ref{fig:ExpSequ_SM}) is used to transfer the spin population from $\ket{\downarrow}$ 
to the $n=58$ hydrogenic manifold via a three-photon transition. 
Atoms in the hydrogenic states (labelled by $\ket{h_i}$ in Fig.\,\ref{fig:ExpSequ_SM}b) are essentially decoupled from those remaining 
in $\ket{\uparrow}$, thus avoiding detrimental effects of interactions during the 
remainder of the read-out sequence.
In the second step, a deexcitation pulse is performed by applying a $2.5~\mu$s laser 
pulse on resonance with the transition between $\ket{\uparrow}$ and the short-lived 
intermediate state $6P_{3/2}$ from which the atoms decay back to $5S_{1/2}$.
The final step consists of switching the tweezers back on to recapture and image 
(via fluorescence) \emph{only} the atoms in $5S_{1/2}$ (while the others are lost). 
Thus we map the $\ket{\uparrow}$ (resp. $\ket{\downarrow}$) state to the 
presence (resp. absence) of the corresponding atom.

\subsection{Experimental imperfections}\label{SM:exp_imperfections}

Several sources of state preparation and measurement (SPAM) error 
contribute to increasing (i.e. worsening) the observed squeezing parameter. \\

\emph{State preparation errors} -- The preparation sequence is composed of two steps: 
the Rydberg excitation (STIRAP) and the  preparation of $\ket{\psi(0)}$ 
by a microwave  $\pi/2$ pulse.
We estimate that the Rydberg excitation process is $98\%$ efficient: 
on average, a fraction $\eta = 2\%$ of the atoms remains in the state $|g\rangle$ 
after excitation and hence do not participate in the dynamics. 
At the end of the sequence, these uninitialized atoms are imaged as a spin $\left\vert \uparrow \right\rangle$.
The $\pi/2$-microwave pulse is also imperfect due to the unavoidable influence of 
the dipolar interactions between the atoms during its application.
Including them in numerical simulations (see Section~\ref{sec:numerics}),
we find that the collective spin state undergoes a slight squeezing dynamics \emph{during} 
the preparation pulse, which reduces the initial polarization by $\sim 1\%$ \\

\emph{Measurement errors} -- 
Due to the finite efficiency of each step in the readout sequence (see Fig.~\ref{SM:Exp_details}b), 
an atom in $\ket{\uparrow}$ (resp. $\ket{\downarrow}$) 
has a non-zero probability $\epsilon_{\uparrow}$ (resp. $\epsilon_{\downarrow}$) 
to be detected in the wrong state~\cite{Chen2023}. 
The main contributions to $\epsilon_{\uparrow}$ are the finite efficiency $1-\eta_{\text{dx}}$ 
of the deexcitation pulse and the probability of loss $\epsilon$ due to collisions with the background gas. 
As for $\epsilon_{\downarrow}$, the main physical origin is the $\ket{\downarrow}$ Rydberg state radiative lifetime.
We use a set of independent measurements and simulations to estimate these imperfections. 
We find to first order $\epsilon_{\uparrow} \simeq \eta_{\text{dx}} + \epsilon = 1.5\% + 1.0\% = 2.5\%$ 
and $\epsilon_{\downarrow} = 1.0\%$.

The finite detection errors impose a lower bound on the observed minimum variance.
More specifically, the experimental magnetizations $\langle J_{y,\theta} \rangle$ and variance 
$\text{Var}(J_\theta)$ are related to the 
same quantities $\langle \tilde{J}_{y,\theta} \rangle$ and $\text{Var}(\tilde{J_\theta})$ 
{\it without} detection errors by the following equations (valid to first order in 
$\epsilon_{\uparrow,\downarrow}$):
\begin{equation}\label{Eq:correction}
\begin{split}
\langle J_{y,\theta} \rangle =~& \frac{N}{2}(\epsilon_{\downarrow}-\epsilon_{\uparrow}) + (1-\epsilon_{\downarrow}-\epsilon_{\uparrow})\langle \tilde{J}_{y,\theta} \rangle\\
\text{Var}\left(J_\theta\right) =~& (1-2\epsilon_{\downarrow}-2\epsilon_{\uparrow})\text{Var}(\tilde{J_\theta}) \\ & + 
\epsilon_{\downarrow}(N/2-\langle \tilde{J_\theta} \rangle)+ 
\epsilon_{\uparrow} (N/2+\langle \tilde{J_\theta} \rangle).
\end{split}
\end{equation}
By inverting the above equations, we calculate the mean-spin length and minimal variance 
free from detection errors (experimentally, the magnetization along the $\theta$-axis, not shown, 
verifies $|\langle J_\theta \rangle|\ll N/2$, leading to a negligible contribution to the correction). 
The  data corrected in this way are shown as diamond symbols in the figures of the main text. 

\subsection{Numerical simulations methods} \label{sec:numerics}

In this section, we provide a summary of the numerical methods used to simulate the experimental 
system and compare experimental findings and theoretical predictions.\\

\emph{Krylov simulations} -- 
For system sizes $2\times 2$, $3\times 3$ and $4 \times 4$, 
numerical simulations were performed with Krylov methods using Dynamite\cite{dynamite}.
Krylov methods are extremely accurate over the short time scales 
relevant to the experiment, so the numerical error associated with this method is negligible.
Moreover, it is straightforward to implement  the aforementioned experimental 
imperfections in these simulations, including missing atoms, 
finite-duration pulses, measurement errors, positional disorder, and van der Waals interactions
(the last two sources of error were described in \cite{Chen2023} and have negligible effect on the squeezing). We note that simulating missing atoms and positional disorder requires significant sampling, which we find converges after $\sim 100$ samples.
Plugging the experimental parameters (interaction strengths, $\eta$ and detection errors) 
into the numerics yields reasonable agreement with the data as shown, 
e.g., in Fig.~\ref{fig:fig2}\textrm{c} and Fig.~\ref{fig:fig3}\textrm{c}.
We attribute the remaining discrepancy to unaccounted-for experimental imperfections -- 
e.g. the effects of other atomic levels outside the $\{\uparrow, \downarrow, \textrm{empty}\}$ manifold, 
decoherence, or microwave control errors 
(resulting in $\sim 3^\circ$ of over/under rotation for $\Omega = 2 \pi \times 4.1$ MHz).\\

\emph{MPO evolution} -- 
For system sizes $6\times 6$ and $8\times 8$, 
we perform numerical simulations using exponential of matrix product operators (MPO) 
as implemented in TenPy \cite{tenpy}. 
In this method, each step of time evolution is implemented as a matrix-product 
operator acting on the matrix product state, which increases the dimension of the matrices.
Then, a truncation is implemented to approximate the quantum state in a 
new matrix-product form with a reduced matrix dimension.
Typically, the accuracy of the method is controlled by the so-called bond dimension 
$\chi$, i.e. the maximum allowable dimension of the matrices in the simulation.
As the entanglement increases during the quantum dynamics, 
a larger bond dimension is required to achieve the same level of accuracy.
In the spin squeezing dynamics at experimentally relevant system sizes, 
the optimal squeezing occurs at early times.
As a result, we find good convergence for bond dimensions ranging from  $\chi=64$ to $\chi=128$. 
We again implement the various imperfections discussed above to obtain 
reasonable agreement with the experimental results.

A larger bond dimension would be required to simulate the $10 \times 10$-atom system 
compared to the $8 \times 8$ one. While the resulting increase in computational memory is affordable, 
the associated increase in computation time is severe. 
Specifically, going from simulating an $8\times 8$ system with $\chi=64$ to a $9\times 9$ system with $\chi=128$ 
increases the computation time from 2-3 days to 8 days. 
This increase is partly due to the larger $\chi$, and partly due to the fact that optimal squeezing occurs at a later time. 
For a $10 \times 10$ system, we estimate a required bond dimension of at least $\chi=198$, 
leading to a simulation time of $\sim 30$ days. 
This corresponds to the time required for simulating a single disorder realization of the Hamiltonian; 
to simulate the experimental error tree, we must sample $\sim 100$ realizations, 
making MPS numerics on the $10 \times 10$ system impractical.

\emph{Time Dependent Variational Monte-Carlo (tVMC)} --
Making use of the time-dependent variational principle, we time-evolve a pair-product state 
(or spin-Jastrow state \cite{Comparin2022PRA}), proven to be extremely 
accurate in describing the dynamics of the dipolar XY model \cite{Comparin2022b}. 
For all system sizes, we simulate the dynamics with open boundary conditions, 
in the ideal case -- \emph{i.e.} we consider the evolution starting from the perfect 
coherent-spin state, and driven exactly by the XY dipolar spin Hamiltonian.

\subsection{Comparison between experimental results and numerical simulations} 

\begin{figure}
\includegraphics{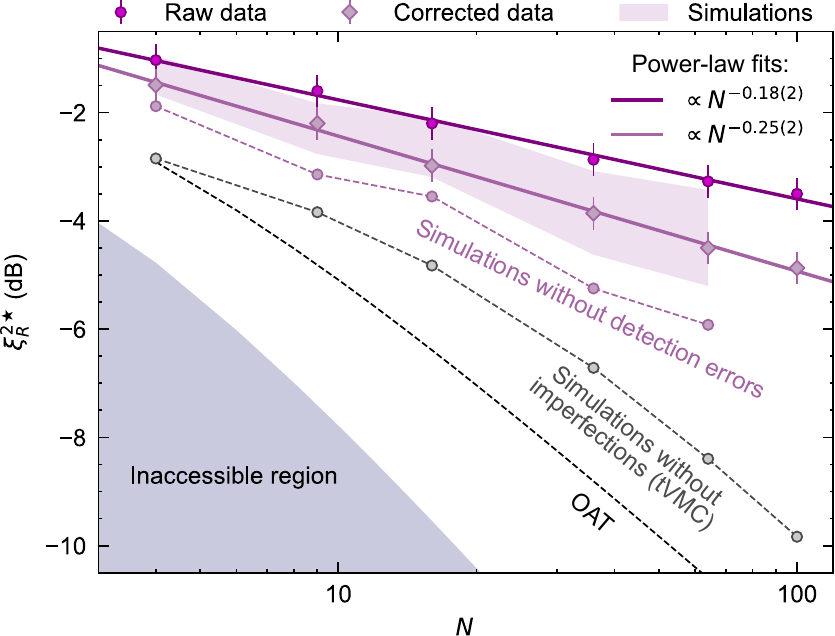}
\caption{\textbf{Minimum squeezing parameter as a function of atom number $N$}.
The circles and diamonds correspond to raw and corrected data, respectively. The solid coloured lines are power-law fits. 
The purple shaded region shows the simulations including $97.5 \pm 1\%$ (resp. $99 \pm 1\%$) detection efficiency of $\ket{\uparrow}$ (resp. $\ket{\downarrow}$). 
The dashed  curves represent the results of simulations of the XY dipolar model 
without state preparation and measurement errors (grey) and without detection errors only (pink). 
The dashed black curve represents the  exact results for the OAT model. 
The inaccessible region corresponds to values of the squeezing parameter smaller than $2/(2+N)$ \cite{Pezze2018}.
}
\label{fig:fig_squeezing_and_simulations}
\end{figure}

The results from all simulations are summarized in Fig.~\ref{fig:fig_squeezing_and_simulations}, 
including various degrees of experimental imperfections. 
The good agreement between the data and the Krylov and MPO simulations including the state preparation and 
measurement errors shows that we 
understand most of the deviations between the experiment and the  perfect dipolar XY dynamics.
The simulations also highlight that even without detection errors, the imperfect state preparation 
contributes to the reduction of the squeezing parameter with respect to the perfect model.  

As mentioned in the main text, although the experiment exhibits scalable squeezing, 
the observed scaling exponent is smaller than expected from the dipolar XY model.
However the results of the simulations for the perfect model indicates that the number of atoms
used in the experiment ($N \leq 100$) only allows us to reach the onset of the predicted asymptotic scaling. 

\subsection{Emergence of OAT-like Squeezing Dynamics in the Dipolar XY model}\label{Sec:XY-OAT}

In this section, we elaborate on the approximation of the $d=2$ dipolar XY model by the OAT model plus ``spin-wave'' corrections (described below).

\emph{Projective Approximation}-- The basis of this approximation is that the dynamics generated by the dipolar XY model, starting from a coherent spin state 
is, \emph{projectively equivalent} to that of the OAT model at short times. 
Indeed the initial coherent spin state lives in the sector of Hilbert space possessing the 
maximal collective-spin of modulus $ {\bf J}^2 = J(J+1)$ with $J = N/2$, which contains 
only permutationally symmetric states, i.e. superpositions of Dicke states. 
Projecting onto this subspace, individual spin operators reduce to collective spin operators, 
e.g. ${\cal P}_{J=N/2} ~\sigma_i^\mu ~{\cal P}_{J=N/2} = 2J_{\mu}/N$  
where $\mu = x,y,z$ and  ${\cal P}_{J=N/2}$ is the projector on the Dicke-state manifold. 
Under the same projection, the dipolar XY Hamiltonian becomes equivalent to the OAT model:
${\cal P}_{J=N/2} H_{XY} {\cal P}_{J=N/2} = J_z^2/(2I) + {\rm const.}$, where 
$1/(2I) = 2J [N(N-1)]^{-1} \sum_{i<j} (a/r_{ij})^3$ ~\cite{Roscilde2023,Roscilde2023b}.
This is nothing but an isolated Anderson tower \cite{anderson_approximate_1952, anderson_basic_1997}, or, equivalently, a quantum rotor with macroscopic spin-length $N/2$ and moment of inertia $I$ (henceforth, ``rotor model'').

The projective equivalence only holds under the assumption that the dipolar dynamics initialized in the Dicke-state manifold remains confined to it. 
This is clearly not the case, since the dipolar Hamiltonian (unlike the OAT) does not conserve ${\bf J}^2$. 
Nonetheless, continuous symmetry breaking at the temperature corresponding to the initial coherent spin state guarantees that the collective spin modulus remains of $O(N^2)$, because the dynamics develops long-range spin-spin correlations. 
This suggests therefore that the evolved state may retain a large overlap with the Dicke manifold. 
Corrections to the projective equivalence picture can be added in the form of spin-wave excitations, which describe the component of the wavefunction leaking into sectors with $J<N/2$.
Such corrections are addressed in the next subsection. 

\begin{figure}[t!]
\includegraphics{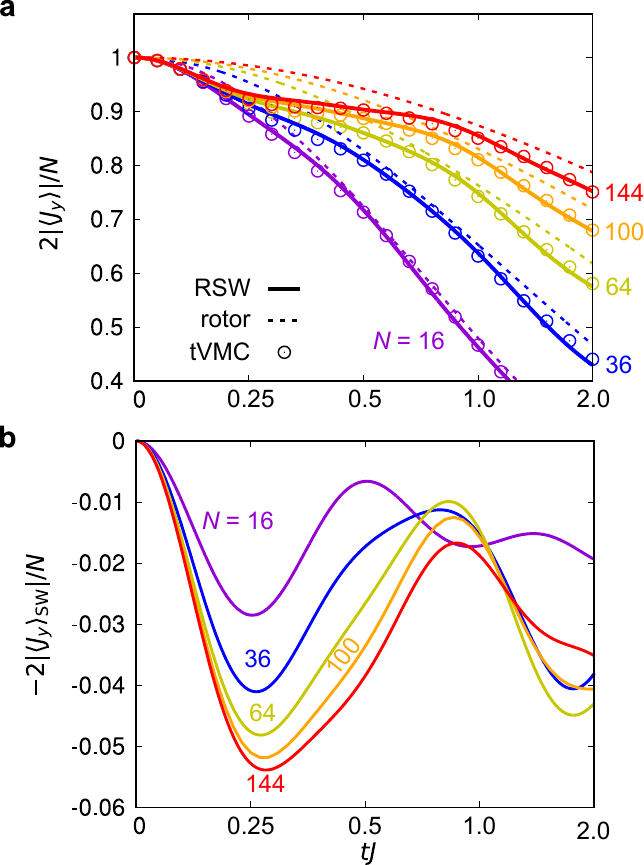}
\caption{\textbf{Magnetization dynamics and its contributions.}
\textbf{a}, Dynamics of the magnetization per spin for the dipolar XY model on a periodic square lattice. 
Results from tVMC calculations and RSW theory for various system sizes ($N=16,..., 144$).
We also show the rotor contribution to the magnetization, corresponding to an effective one-axis-twisting model (see text). 
\textbf{b}, Spin-wave (SW) contribution to the magnetization.}
\label{fig:MagnDyn_SM}
\end{figure}

\emph{Spin-wave Corrections}-- Here, we provide further insight on the demagnetization dynamics and its scaling properties with the atom number $N$, as observed in Fig.\,\ref{fig:fig3}a. It relies on the rotor/spin-wave (RSW) theory \cite{Roscilde2023,Roscilde2023b}, which allows one to write the magnetization as $\langle J_y\rangle = \langle J_y\rangle_{\rm R} + \langle J_y\rangle_{\rm SW}$.
Here, $\langle J_y\rangle_{\rm R}$ is the magnetization of the macroscopic spin of length $N/2$ (the rotor) introduced in the above. 
It obeys the dynamics of the one-axis-twisting model $(J_z)^2/(2I)$; 
$\langle J_y\rangle_{\rm SW}$ is a (negative) spin-wave (SW) contribution, 
coming from linear excitations at finite momentum which are triggered by the quantum quench dynamics. 

RSW theory quantitatively accounts for the magnetization dynamics of systems with periodic boundary conditions, as shown by the excellent agreement with the tVMC results (see Fig.~\ref{fig:MagnDyn_SM}a) \cite{Roscilde2023}.
Owing to the very low density of spin-wave excitations triggered by the dipolar XY dynamics initialized in the coherent spin state, the validity of RSW theory stretches to rather long times, well beyond those explored in our experiment.
In particular, RSW theory elucidates the scaling properties of the magnetization dynamics at short times. 
As shown in the experimental data of Fig.\,\ref{fig:fig3}a, as well as in the simulations in Fig.~\ref{fig:MagnDyn_SM}a, the magnetization per spin exhibits an initial decay  independent of system size up to a time scale $t_{\rm SW} \sim 1/(4J)$. 
On the contrary, the later dynamics acquire a strong system-size dependence. 
RSW theory indicates that the initial size-independent decay comes from the proliferation of spin-wave excitations (appearing in counter-propagating pairs) at a time scale $t_{\rm SW}$ 
indicated by the first local minimum in Fig.~\ref{fig:MagnDyn_SM}b, 
marking the saturation of the spin-wave population to its first maximum.
By contrast, the later, size-dependent dynamics is dominated by the rotor variable, 
which depolarizes as in the one-axis-twisting model, namely over time scales growing as $\sqrt{N}$. 
The longer persistence of magnetization for larger system sizes is the finite-size precursor of spontaneous symmetry breaking of the U(1) symmetry, which appears in the thermodynamic limit for both the one-axis twisting and dipolar XY models.

\subsection{Multi-step Squeezing}

\begin{figure*}
\centering\includegraphics[width=\textwidth]{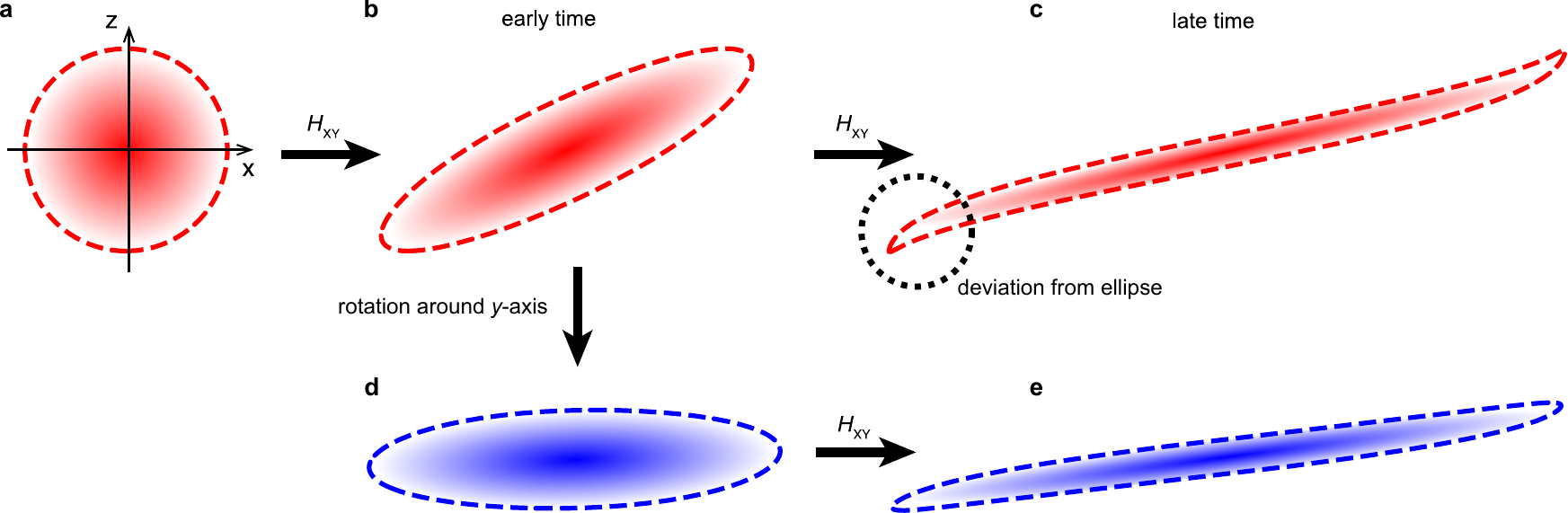}
\caption{\textbf{Schematic depicting the multi-step squeezing protocol.} 
\textbf{a}, Semi-classical description of a y-polarized initial state. 
\textbf{b}, \textbf{c}, Normal spin squeezing dynamics. 
\textbf{d}, \textbf{e}, Multi-step squeezing dynamics enabled by an extra rotation along the mean spin direction. }
\label{fig:multi_schematic}
\end{figure*}

In the main text, we implement the multi-step squeezing protocol and demonstrate that it improves the optimal squeezing. 
Here, we provide further analysis from a theoretical perspective, as well as comparison between numerical simulation and experiments. \\

\emph{Physical intuition} --
The squeezing dynamics can be understood within a semi-classical picture \cite{Block2023}: 
treating the total spin as an ensemble of classical points, 
the initial state $\ket{\psi(0)}$ is represented by a Gaussian distribution with the same variance 
$\sim\sqrt{N}$ along the $z$ and $x$ axes (i.e. a disc in the $x-z$ plane); 
in the dynamics, each  point rotates around the $z$-axis with an angular 
velocity proportional to its $z$ polarization. 
The corresponding classical equations of motion are:
\begin{equation}
\begin{split}
x(t) &= x(0)+N m_{\mathrm{xy}}\,\mathrm{sin}\left(\frac{\tilde{J} z t}{N}\right), \\
z(t) &= z(0), 
\end{split}
\end{equation}
where $\tilde{J}$ and $m_{xy}$ are the effective coupling strength and the effective total spin length. 
Consequently, the circle approximately deforms into an ellipse in the $x-z$ plane.
However, the ellipse is perfect only for small $z$ and short times, i.e. $\tilde{J} z t\ll N$ and thus 
$N\mathrm{sin}(\tilde{J} z t/N)\approx \tilde{J} z t$ [Fig.~\ref{fig:multi_schematic}(b)]. 
If such a condition were always satisfied, the squeezing parameter would keep improving for all time. 
Instead, the optimal squeezing is achieved when the deviation from a perfect ellipse 
(which happens at earlier times for larger $z$) becomes larger than the minor-axis of the ellipse [Fig.~\ref{fig:multi_schematic}(c)]. 
Therefore, a natural way to improve squeezing is to delay the time when such a deviation happens: 
before the deviation becomes the bottleneck, one can rotate the major axis of the ellipse towards 
$x$-axis [Fig.~\ref{fig:multi_schematic}(d)], so that the typical $z$ value of the classical ensemble 
becomes smaller, delaying the non-elliptical deviation to later times [Fig.~\ref{fig:multi_schematic}(e)]. \\

\begin{figure}
\includegraphics{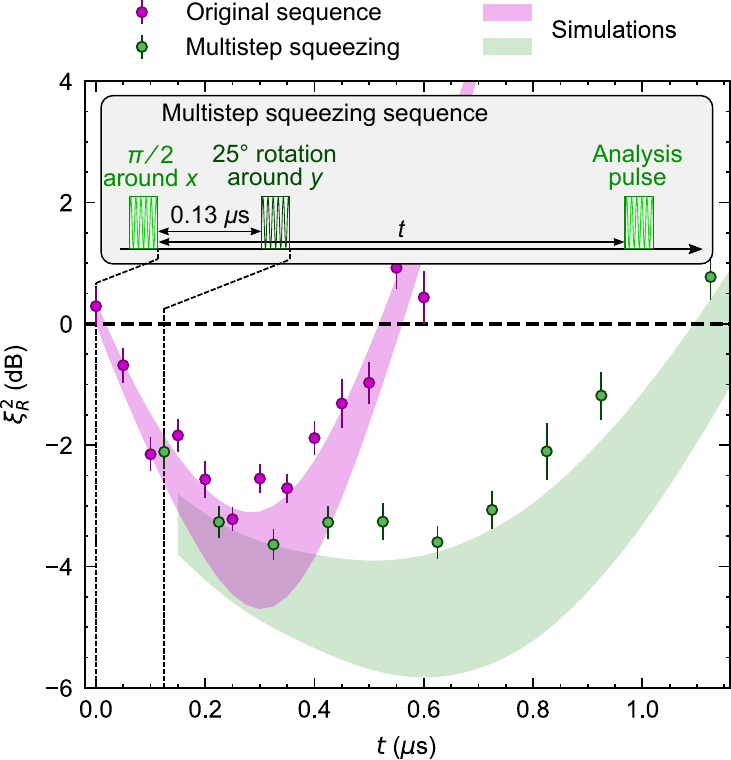}
\caption{\textbf{Multistep squeezing, comparison between data and simulation}
Measurements of the squeezing parameter obtained with two different procedures. The first one (purple dots) is the original sequence illustrated in  Fig.~\ref{fig:fig1}(c). The second one (dark green dots) is the multistep sequence. The shaded regions show the simulations including $97.5 \pm 1\%$ (resp. $99 \pm 1\%$) detection efficiency of $\ket{\uparrow}$ (resp. $\ket{\downarrow}$). These data correspond to a $6\times6$ array.
}
\label{fig:fig_SM_multistep_squeezing}
\end{figure}

\emph{Comparison between numerics and
experimental data} -- 
Similar to the single-step squeezing, we also performed TEBD simulation for multi-step squeezing dynamics, 
taking all experimental imperfections into account. The results are shown in Fig.~\ref{fig:fig_SM_multistep_squeezing}, 
where observe a relatively good agreement between the numerics and the experimental data. 

\end{document}